\documentclass[prl,twocolumn,showpacs,superscriptaddress]{revtex4-1}
\usepackage{amsmath,graphicx}

\usepackage{mathptmx}

\newcount\minute
\newcount\hour
\newcount\hourMins
\def\now
{
  \minute=\time
  \hour=\time \divide \hour by 60
  \hourMins=\hour \multiply\hourMins by 60
  \advance\minute by -\hourMins
  \zeroPadTwo{\the\hour}:\zeroPadTwo{\the\minute}
}\def\timestamp
{
  \today\ \now
}
\def\today
{
  \zeroPadTwo{\the\day}-\zeroPadTwo{\the\month}-\the\year%
}
\def\zeroPadTwo#1%
{
  \ifnum #1<10 0\fi
  #1%
}
\def \dif {{d}}

\date{\timestamp}

\begin{document}

\title{Decay of Bogoliubov quasiparticles in a nonideal one-dimensional Bose gas}

\author{Zoran Ristivojevic}
\affiliation{Centre de Physique Th\'{e}orique, Ecole Polytechnique, CNRS, 91128 Palaiseau, France}

\author{K. A. Matveev}
\affiliation{Materials Science Division, Argonne National Laboratory, Argonne, Illinois 60439, USA}

\begin{abstract}
We study the relaxation of excitations in a system of one-dimensional weakly interacting bosons. Due to residual weak interactions, Bogoliubov quasiparticles in this system have finite lifetimes. As a result of the conservation laws in one dimension, at zero temperature the leading mechanism of decay of a quasiparticle is disintegration into three others. We focus on phonon quasiparticles and find that their decay rate is proportional to the seventh power of momentum. In the integrable case of contact interaction between the bosons, the decay rate vanishes.
\end{abstract}

\maketitle

The excitations of a three-dimensional system of interacting bosons are Bogoliubov quasiparticles \cite{bogoliubov47}. At low energies these quasiparticles are phonons with linear spectrum. Due to residual interactions the quasiparticles have finite lifetimes. At zero temperature, the leading decay process, known as Beliaev decay, involves disintegration of a single quasiparticle into two others. The resulting decay rate of a phonon is proportional to the fifth power of its momentum \cite{Beliaev58,landaubook-9}. The Beliaev theory is experimentally confirmed in three-dimensional Bose-Einstein condensates  \cite{hodby+01PhysRevLett.86.2196, katz+02-beliaevdamping3d-PhysRevLett.89.220401}.

A great deal of attention devoted to low-dimensional bosonic systems is boosted by recent experimental progress with ultracold atoms \cite{kinoshita+04tonksgas,paredes+04tonksgas, hadzibabic+2006berezinskii,bloch+08RevModPhys,langen+13}. Nevertheless, understanding the nontrivial effects of interaction in one-dimensional bosonic systems  remains a challenge. In particular, the experimental study of an initially prepared Bose gas in a nonequilibrium state shows the absence of equilibration of excitations \cite{kinoshita+06equilibration}. There the authors studied the systems of bosons with a short-range interaction. Theoretically, in the limit of contact interaction, this system is described by the Lieb-Liniger model \cite{lieb1963exact}, which is integrable, and therefore there should be no damping of excitations \cite{kinoshita+06equilibration}.

One-dimensional interacting bosons are conventionally treated in terms of the Luttinger liquid theory \cite{haldane81prl,cazalilla+2011RMP}. The excitations of a Luttinger liquid are noninteracting phonons with infinite lifetimes. Even after one amends the theory with anharmonic corrections that account for the interactions between the excitations, the evaluation of their lifetimes is a challenging problem \cite{imambekov+12RevModPhys.84.1253}. This is because the excitations of the Luttinger liquid have a linear dispersion, and therefore for phonons on the same branch conservation of momentum guarantees conservation of energy. The resulting infinite degeneracy of the multi-phonon states gives rise to singularities in the perturbation theory for the phonon decay rate. An attempt to deal with this problem was made in Ref.~\cite{samokhin98} by using a self-consistent approximation \cite{andreev80} and resulted in a decay rate that scales as the square of the momentum. Interestingly, this result persists even in integrable models, in an apparent contradiction with Ref.~\cite{kinoshita+06equilibration}.

The origin of this discrepancy is that phonons are not the true quasiparticles of the system at lowest momenta. Instead, at $q\to 0$, the quasiparticles of a generic one-dimensional quantum liquid are fermions \cite{rozhkov2005fermionic,imambekov+12RevModPhys.84.1253}. Their spectrum is given by
\begin{align}\label{fermionic spectrum}
\epsilon_q=v|q|+{q^2}/{2m^*},
\end{align}
where $v$ is the sound velocity in the system, and the effective mass $m^*$ accounts for the nonlinearity arising from the leading irrelevant perturbation in the Luttinger liquid. The remaining perturbations lead to scattering of fermionic quasiparticles. The resulting decay rate of a quasiparticle is proportional to the eighth power of momentum
\cite{khodas+07PhysRevB.76.155402,matveevfurusaki13}. Thus, at $q\to 0$, the decay of fermionic quasiparticles is much slower than that of phonons in Ref.~\cite{samokhin98}.

In this article we consider the problem of decay of excitations in a one-dimensional Bose gas with weak short-range repulsion. Similarly to higher dimensions, the excitations have bosonic statistics and satisfy the Bogoliubov dispersion relation \cite{kulish+1976comparison}
\begin{align}\label{bogoliubov spectrum}
\varepsilon_q=v|q|\sqrt{1+2q^2/q_0^2},\quad q_0=\sqrt{8}mv.
\end{align}
Here $m$ denotes the mass of the physical particles forming the Bose gas. The phonon part of the excitation spectrum, $q\ll q_0$, is approximately linear, and the system can be treated in the framework of the Luttinger liquid theory. It is important to note, however, that the dispersion (\ref{bogoliubov spectrum}) does have a finite curvature. The cubic nonlinearity in Eq.~(\ref{bogoliubov spectrum}) is comparable to the quadratic nonlinearity of the effective fermionic excitations in Eq.~(\ref{fermionic spectrum}) at momentum $q^*\sim (m/m^*)q_0$. At $q\ll q^*$ the curvature of the spectrum (\ref{bogoliubov spectrum}) is small and can be accounted for as an additional perturbation in the picture of fermionic quasiparticles. Conversely, at $q\gg q^*$, the curvature of the bosonic spectrum (\ref{bogoliubov spectrum}) is the leading correction to the Luttinger liquid Hamiltonian. In this regime the phonons are the true quasiparticles of the system \cite{pustilnik+14}. The curvature lifts the degeneracy of the multi-phonon states that resulted in singularities encountered in Refs.~\cite{samokhin98,andreev80}, and the scattering of phonons can be studied perturbatively \cite{Lin+13PhysRevLett.110.016401}.

The interaction strength in a Luttinger liquid is quantified by the parameter $K=\pi\hbar n_0/mv$, where $n_0$ is the mean particle density. For weakly interacting bosons, $K\gg 1$. This regime is routinely realized in experiments. For example, the value $K=42$ was reported in Ref.~\cite{hofferberth+08probing}, while the particle density range explored in the experiment \cite{kruger+10PhysRevLett.105.265302} corresponds to $9<K< 84$. It is important to note that at weak interactions, the crossover momentum $q^*\sim (m/m^*)q_0$ is small, $q^*\ll q_0$. Indeed, the effective mass of fermionic quasiparticles is related to the mass of physical particles as $m^*=4m\sqrt{K}/3$ \cite{imambekov+12RevModPhys.84.1253, pereira+2006PRL}. Therefore, at $K\gg 1$, the phonons exist in a broad range of momenta between $q^*\sim q_0/\sqrt{K}$ and $q_0$. Our main goal is to study their decay. Instead of relying on the phenomenological approach based on the Luttinger liquid theory, we take advantage of the weak interaction strength and develop a microscopic theory. At $q\gg q^*$, we find results that are different from those of Ref.~\cite{samokhin98}.

We start by considering the kinetic energy of bosons $({\hbar^2}/{2m})\int\dif x (\nabla\Psi^\dagger)(\nabla\Psi)$, where $\Psi(x)$ and $\Psi^\dagger(x)$ are the bosonic single particle operators. After introducing the representation
$\Psi^\dagger=\sqrt n\,e^{i\theta}$ \cite{popov72,haldane81prl}, where the density $n$ and the phase $\theta$ satisfy the standard bosonic commutation relation $[n(x),\theta(y)]=-i\delta(x-y)$, the kinetic energy takes the form
\cite{popov72}
\begin{align}\label{H}
H_{\mathrm{kin}}=\frac{\hbar^2}{2m}\int\dif x\left[n(\nabla\theta)^2+\frac{(\nabla n)^2}{4n}\right].
\end{align}

The effects of interaction in our system of one-dimensional bosons are described by
\begin{align}\label{H'}
H_{\mathrm{int}}=\frac{g}{2}\int\dif x\,n^2-\frac{\hbar^2}{m}\alpha\int\dif x\,n^3.
\end{align}
The first term in Eq.~(\ref{H'}) accounts for contact repulsion between the bosons. The parameter $g>0$ describes the interaction strength. In order to obtain a nonvanishing decay rate, we included the second term in Eq.~(\ref{H'}). It has the form of a three-body interaction and represents the leading integrability breaking perturbation \cite{muryshev+02PhysRevLett.89.110401, mazets+08PhysRevLett.100.210403,tan+10relaxation}, with the dimensionless parameter $\alpha$ characterizing its strength \footnote{The second term in $H_{\mathrm{int}}$ is a consequence of the particular experimental realization of one-dimensional bosons, obtained by tight confinement of three-dimensional particles \cite{mazets+08PhysRevLett.100.210403}. Another way to break integrability is by a non-contact short-range interaction, which leads to qualitatively similar conclusions. In particular, the decay rate (\ref{decay T=0}) has the same $Q^7$ behavior at small momentum.}.

We treat the Hamiltonian $H=H_{\mathrm{kin}}+H_{\mathrm{int}}$ using the standard procedure \cite{haldane81prl,cazalilla+2011RMP} to account for small density fluctuations. We express the bosonic density operator as
\begin{align}\label{density}
n=n_0+\nabla\varphi/\pi,
\end{align}
where the field $\varphi$ is defined by the commutation relation $[\nabla\varphi(x),\theta(y)]=-i\pi\delta(x-y)$. In a theory limited to the excitations of wave vectors smaller than $n_0$, the fluctuations of the field $\nabla\varphi$ are small. This enables us to expand the Hamiltonian (\ref{H}) in powers of $\nabla\varphi$ and then to collect terms with different powers of the bosonic fields $\theta$ and $\varphi$. We start with the quadratic terms. The resulting quadratic Hamiltonian can be diagonalized using the following relations to connect the bosonic fields $\varphi$ and $\theta$ with the bosonic quasiparticle operators $b_q$ and $b_q^\dagger$,
\begin{gather}\label{nablaphi}
\nabla\varphi(x)=\sum_q \sqrt{\frac{\pi^2 n_0}{2Lm\varepsilon_q}}|q|e^{i q x/\hbar} (b_{-q}^\dagger+b_q),\\
\label{nablatheta}
\nabla\theta(x)=\sum_q \sqrt{\frac{m \varepsilon_q}{2 L\hbar^2 n_0}}\,\text{sgn}(q)e^{i q x/\hbar} (b^\dagger_{-q}-b_{q}).
\end{gather}
Here $L$ is the system size. Upon diagonalization, the quadratic Hamiltonian takes the form
\begin{align}\label{H2}
H_{0}=\sum_q\varepsilon_q b_q^\dagger b_q,
\end{align}
where the excitation spectrum is given by the Bogoliubov dispersion (\ref{bogoliubov spectrum}) with the sound velocity
\begin{align}\label{v}
v=\sqrt{gn_0/m-6\alpha\hbar^2n_0^2/m^2}.
\end{align}

In addition to the quadratic terms contained in the Hamiltonian (\ref{H2}), there are higher order terms that account for the interaction between excitations. The cubic in $\varphi$ and $\theta$ correction to $H_0$ reads
\begin{align}\label{H3general}
V_3=&\frac{\hbar^2}{m}\int\dif x \biggl[\frac{1}{2\pi} (\nabla\varphi)(\nabla\theta)^2- \frac{1}{8\pi^3n_0^2}(\nabla^2\varphi)^2(\nabla\varphi)\notag\\
  &-\frac{\alpha}{\pi^3}(\nabla\varphi)^3 \biggr].
\end{align}
The first term in Eq.~(\ref{H3general}) arises from the first term in the kinetic energy (\ref{H}). It has a scaling dimension three and represents the leading perturbation to the Hamiltonian (\ref{H2}). The second term in Eq.~(\ref{H3general}) originates from the second term in Eq.~(\ref{H}), the so-called quantum pressure. It has a scaling dimension five and thus is commonly neglected in the standard theory of interacting bosons \cite{cazalilla+2011RMP}. However, we will see below that despite being of a higher scaling dimension than the first term in Eq.~(\ref{H3general}), it must be included in a consistent theory of quasiparticle decay. The last term in Eq.~(\ref{H3general}) is of a scaling dimension three and arises from the second term in Eq.~(\ref{H'}).

In addition to the terms included in Eq.~(\ref{H3general}), in a phenomenological approach one would expect to find further corrections to the Hamiltonian (\ref{H2}). In order to obtain the leading order result for the decay rate of phonon excitations, certain quartic terms would have to be included, such as the one proportional to $(\nabla\varphi)^4$. However, in our microscopic theory of weakly interacting bosons described by Eqs.~(\ref{H}) and (\ref{H'}), such terms do not appear.

\begin{figure}
\includegraphics[width=0.5\columnwidth]{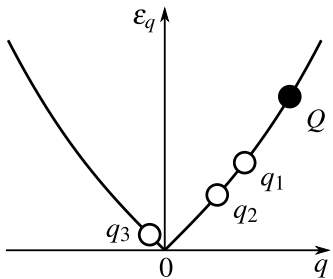}
\caption{In a one-dimensional Bose gas, a phonon excitation of momentum $Q$ decays into three excitations with momenta $q_1,q_2$, and $q_3$. Using the conservation laws, one finds that two phonons in the final state propagate in the direction of the initial phonon, while the remaining one is counterpropagating.}\label{fig1}
\end{figure}

For the curved spectrum (\ref{bogoliubov spectrum}), conservation laws do not allow a phonon to decay into two others. The leading decay process is disintegration of a phonon into three other phonons (Fig.~\ref{fig1}). At zero temperature, the decay rate for this process is determined by the Fermi golden rule,
\begin{align}\label{rate def}
\frac{1}{\tau}=\frac{2\pi}{\hbar} \sum_{q_1>q_2>q_3}\!\!|\mathcal{A}_{Q}^{q_1,q_2,q_3}|^2 \delta(\varepsilon_{Q}-\varepsilon_{q_1}- \varepsilon_{q_2}-\varepsilon_{q_3}).
\end{align}
Here the $\delta$ function accounts for the conservation of energy. The matrix element $\mathcal{A}_{Q}^{q_1,q_2,q_3}$ describes the decay of the initial quasiparticle excitation of momentum $Q$ into three phonons with momenta $q_1$, $q_2$, and $q_3$. It is defined in terms of the $T$-matrix as $\mathcal{A}_{Q}^{q_1,q_2,q_3} =\langle 0| b_{q_1} b_{q_2} b_{q_3}|T|b^\dagger_{Q}|0\rangle$.

The dominant contribution to the scattering matrix element arises in the second order in perturbation (\ref{H3general}). Because $\nabla\varphi$ and $\nabla\theta$ in Eq.~(\ref{H3general}) enter the expression for the amplitude via the normal mode expansions (\ref{nablaphi}) and (\ref{nablatheta}), each of the creation and annihilation operators $b_q$ and $b_q^\dagger$ is accompanied by a factor of $\sqrt{|q|}$. It is thus convenient to express the scattering amplitude as
\begin{align}\label{amplitude}
\mathcal{A}_{Q}^{q_1,q_2,q_3}= \frac{\Lambda(Q,q_1)}{2L n_0m} \sqrt{|Qq_1q_2q_3|}\,\delta_{Q,q_1+q_2+q_3}.
\end{align}
Here the Kronecker delta reflects the momentum conservation, while the dimensionless function $\Lambda$ depends on only two momenta, $Q$ and $q_1$, because the values of the other two are fixed by the conservation laws.

When $q_1$ approaches $Q$ the momenta of the other two phonons are small,  $q_2,q_3\ll Q$. In this limit one can employ the mobile impurity formalism \cite{landau+49,khodas+07PhysRevB.76.155402, imambekov+12RevModPhys.84.1253,matveev+12PhysRevB.86.045136, schecter+12} to find $\Lambda$ in Eq.~(\ref{amplitude}). Within this approach one treats the initial excitation of momentum $Q$ as a mobile impurity interacting with the host system. Consider the process where the impurity of momentum $Q$ scatters into $Q-\delta Q$, at the same time creating two excitations of momenta $q_2$ and $q_3$ in the host system. Here $\delta Q=Q-q_1$ is small compared to $Q$. The scattering matrix element for such a process is given by Eqs.~(49) and (54) of Ref.~\cite{matveev+12PhysRevB.86.045136}. It is expressed in terms of the impurity spectrum and its dependence on the density of the host liquid. Substituting the Bogoliubov dispersion (\ref{bogoliubov spectrum}) for the mobile impurity, at $Q\ll q_0$ we find an expression consistent with Eq.~(\ref{amplitude}), provided
\begin{align}\label{C mobile impurity}
\Lambda=\frac{2n_0^2}{v^2}\frac{\dif}{\dif n_0} \left(\frac{v^2}{n_0}\right) + \frac{n_0^3}{2v^2}\frac{\dif^2}{\dif n_0^2} \left(\frac{v^2}{n_0}\right).
\end{align}
For the Lieb-Liniger model we have $\alpha=0$. In this case,  $v^2\propto n_0$ [see Eq.~(\ref{v})], and thus $\Lambda=0$. This is in accordance with the expectation that excitations in integrable models do not decay. The presence of the integrability breaking perturbation (\ref{H'}) affects the sound velocity (\ref{v}) resulting in
\begin{align}\label{C}
\Lambda=-12{\alpha}K^2/{\pi^2}.
\end{align}
We note that Eq.~(\ref{C mobile impurity}) is valid for any dependence of the velocity on density. In particular, it can account for perturbations to the Hamiltonian that have the form of an arbitrary function of density.

The dimensionless quantity $\Lambda$ in Eq.~(\ref{amplitude}) depends on two momenta, $Q$ and $q_1$. The preceding discussion based on the treatment of the phonon $Q$ as a mobile impurity relies on the smallness of the momentum change $\delta Q$. Thus our results (\ref{C mobile impurity})
and (\ref{C}) give the value of $\Lambda(Q,q_1)$ at $q_1=Q$.  It is important to note that Eq.~(\ref{C mobile impurity}) obtained in the limit $Q\ll q_0$ is independent of $Q$. Because of this lack of scaling at small $Q$, one may expect that at low momenta $\Lambda$ approaches the value
(\ref{C mobile impurity}) for any ratio $q_1/Q$.  This conjecture is supported by the full microscopic calculation of the scattering matrix element, which we outline below.

The perturbation (\ref{H3general}) to the quadratic Hamiltonian (\ref{H2}) contains three terms.  At the first step we neglect the second term in Eq.~(\ref{H3general}), as it has a higher scaling dimension.  Because of the near degeneracy of the slightly curved Bogoliubov spectrum in Fig.~\ref{fig1}, at $Q\to 0$ some of the energy denominators in the second-order perturbation theory expression for $\Lambda$ scale as fast as $Q^3$, whereas the numerator scales only linearly with $Q$.  One may therefore expect the leading contribution to $\Lambda$ to scale as $1/Q^2$, which would contradict Eq.~(\ref{C mobile impurity}).  In reality these leading order terms cancel, and one has to account for subleading contributions.  Thus a consistent microscopic theory must include the second term in Eq.~(\ref{H3general}) despite its higher scaling dimension.  A careful calculation \cite{ristivojevic+matveev-unpublished}
recovers the momentum-independent result (\ref{C}).

We are now in a position to calculate the decay rate (\ref{rate def}). The scaling of $1/\tau$ with the momentum of the initial phonon $Q$ can be understood as follows. Conservation laws require that two out of three phonons in the final state propagate in the same direction as the initial phonon, while the third one is on the opposite branch (see Fig.~\ref{fig1}). The momentum $q_3$ of the latter phonon is controlled by the curvature of the spectrum and scales as $Q^3$, whereas the momenta of the other two phonons in the final state, $q_1$ and $q_2$, are of the order of $Q$. Since the value for $\Lambda$ is given by the momentum independent expression (\ref{C}), the square of the matrix element (\ref{amplitude}) scales as $Q^6$. Due to the conservation laws, the phase space volume for scattering in Eq.~(\ref{rate def}) is linear in $Q$, resulting in the decay rate of the initial phonon proportional to the seventh power of momentum. Indeed, substitution of the amplitude (\ref{amplitude}) in the expression for the decay rate (\ref{rate def}) yields
\begin{align}\label{decay T=0}
\frac{1}{\tau}=\frac{144\sqrt{2}}{5\pi}\alpha^2 \frac{T_d}{\hbar}\left(\frac{Q}{q_0}\right)^7.
\end{align}
Here $T_d=\hbar^2 n_0^2/m$ denotes the quantum degeneracy temperature, while the parameter $\alpha$ is defined in Eq.~(\ref{H'}). Expression (\ref{decay T=0}) is our main result.

The decay rate (\ref{decay T=0}) can be contrasted with the result for the three-dimensional Bose gas where $1/\tau\propto Q^5$ \cite{Beliaev58} and with the result $1/\tau\propto Q^8$ for the fermionic quasiparticles in the one-dimensional Bose gas at $Q\ll q^*$ \cite{matveevfurusaki13}. Expression (\ref{decay T=0}) applies to excitations with momenta in the range $q^*\ll Q\ll q_0$. At very high momenta $Q\gg q_0$, the spectrum (\ref{bogoliubov spectrum}) is quadratic. The decay of excitations in this limit was recently studied in Refs.~\cite{tan+10relaxation, mazets+08PhysRevLett.100.210403}. Our microscopic theory can be extended to describe the crossover between these two regimes \cite{ristivojevic+matveev-unpublished}.

Our result (\ref{decay T=0}) is derived at zero temperature. Finite temperature $T$ does not significantly affect this decay rate as long as it is smaller than the typical energy $vQ^3/q_0^2$ of the counterpropagating phonon. One can easily show that at $vQ^3/q_0^2\ll T\ll vQ$, the decay rate (\ref{decay T=0}) becomes
\begin{align}
\frac{1}{\tau}=\frac{96\sqrt{2}}{\pi}\alpha^2 \frac{T_d}{\hbar}\frac{TQ^4}{v q_0^5}.
\end{align}
Finally, for thermal phonons, $vQ\sim T$, the typical decay rate behaves as $T^5$. In this regime, the decay rate cannot be characterized by a unique expression, but rather by a whole spectrum, in analogy to the relaxation of phonons in a one-dimensional Wigner crystal \cite{Lin+13PhysRevLett.110.016401}.

In conclusion, we have studied the intrinsic damping of Bogoliubov quasiparticles in a system of weakly interacting bosons in one dimension. We found that the leading mechanism is decay of a phonon into three other phonons. This is in contrast with the Beliaev decay in two- and three-dimensional systems where only two quasiparticles are present in the final state. At zero temperature, we found the resulting decay rate to be proportional to the seventh power of momentum. Our main result (\ref{decay T=0}) gives the width of the peak in the dynamic structure factor \cite{khodas+07bosonsPhysRevLett.99.110405} that can be measured experimentally \cite{fabbri+11PhysRevA.83.031604}.

We acknowledge stimulating discussions with L.~I.~Glazman and M. Pustilnik. Z.R.~acknowledges the hospitality of INT of the University of Washington, Seattle, where this work began. Work by Z.R.~was supported by PALM Labex. Work by K.A.M.~was supported by the U.S.~Department of Energy, Office of Science, Materials Sciences and Engineering Division.

%

\end{document}